**Fluorescence thermometry enhanced by the quantum coherence of single spins in diamond**

David M. Toyli[a], Charles F. de las Casas[a,1], David J. Christle[a,1], Viatcheslav V. Dobrovitski[b], and David D. Awschalom[a,c,2]

[a]Center for Spintronics and Quantum Computation, University of California, Santa Barbara, California 93106, United States
[b]Ames Laboratory, US Department of Energy, Ames, Iowa 50011, United States
[c]Institute for Molecular Engineering, University of Chicago, Chicago, Illinois 60637, United States

[1]Present address: Institute for Molecular Engineering, University of Chicago, Chicago, Illinois 60637, United States
[2]To whom correspondence should be addressed: awsch@uchicago.edu

**We demonstrate fluorescence thermometry techniques with sensitivities approaching 10 mK Hz$^{-1/2}$ based on the spin-dependent photoluminescence of nitrogen vacancy (NV) centers in diamond. These techniques use dynamical decoupling protocols to convert thermally induced shifts in the NV center's spin resonance frequencies into large changes in its fluorescence. By mitigating interactions with nearby nuclear spins and facilitating selective thermal measurements, these protocols enhance the spin coherence times accessible for thermometry by 45x, corresponding to a 7x improvement in the NV center's temperature sensitivity. Moreover, we demonstrate these techniques can be applied over a broad temperature range and in both finite and near-zero magnetic field environments. This versatility suggests that the quantum coherence of single spins could be practically leveraged for sensitive thermometry in a wide variety of biological and microscale systems.**

Thermometry based on thermally driven changes in fluorescence lifetimes or intensities are essential techniques in many environments that preclude electrical probes (1). Although typical fluorescence thermometers utilize millimeter-scale optical probes (2), the desire to noninvasively monitor intracellular thermal gradients has motivated efforts to develop analogous methods at the nanoscale (3, 4). This interest has stimulated the development of nanoscale fluorescence thermometers based on quantum dots (5), rare-earth ions (6), and nanogels (7), with recent studies suggesting that intracellular temperature gradients are on the order of 1 K (8). However, the application of conventional fluorescence thermometry techniques in biological settings is limited by temperature resolutions of ~0.2 K or worse (8-10), motivating the development of more advanced nanoscale thermometers.

In recent years solid-state electronic spins have gained considerable attention for applications in nanoscale sensing. In particular, the diamond nitrogen-vacancy (NV) center (Fig. 1A) has garnered attention for its optical spin initialization and fluorescence-based spin readout (11), the ability to isolate and measure single defects (12), and the ability to manipulate its spin using microwave electron spin resonance techniques (13). NV center sensing is based on monitoring shifts in the spin resonance frequencies through the defect's fluorescence as a function of external perturbations such as magnetic fields (14-17), electric fields (18), or temperature (19, 20), with the sensitivity of these techniques scaling as $1/\sqrt{T_C}$, where $T_C$ is the relevant spin coherence time (21). These coherence times can be enhanced by 2-3 orders of magnitude for perturbations amenable to AC modulation through the use of dynamical decoupling techniques that periodically invert the spin state and the signal being sensed in order to mitigate the effects of low-frequency magnetic noise (22-24). These methods have enabled the detection of single electronic spins (15) and nanoscale nuclear spin ensembles external to the diamond

(16, 17), with motivations to extend these techniques to biological systems using NV centers in nanodiamonds (25).

Here we demonstrate that the quantum coherence of a single NV center can enhance thermometry. Our technique relies on the use of dynamical decoupling protocols that manipulate the NV center's spin-one ground state to maintain DC sensitivity to thermally driven shifts in the spin resonance frequencies while suppressing the effects of low frequency magnetic noise. This allows us to exploit the NV center's long spin coherence for thermometry. We achieve spin coherence times for thermal sensing exceeding 80 µs, resulting in inferred thermal sensitivities of 25 mK Hz$^{-1/2}$. By combining these prolonged coherence times with improved fluorescence detection (26), we estimate that thermal sensitivities better than 10 mK Hz$^{-1/2}$ should be achievable. These sensitivities demonstrate the NV center's potential to markedly improve on the ~0.2 K temperature resolution offered by competing nanoscale fluorescence thermometry techniques (3, 4, 8-10). We execute these measurements in two regimes: in moderate magnetic fields where both spin transitions can be resonantly addressed with separate microwave signals and near zero magnetic field where they can be addressed with a single microwave frequency. Finally, we demonstrate the wide range of operating temperatures for NV center thermometers by performing these measurements at 500 K. Together, these results provide a robust set of protocols for NV center based thermometry.

**Principles of Spin-Based Fluorescence Thermometry**

Our work focuses on the precise measurement of the NV center's temperature-dependent crystal field splitting ($D$) between its $m_S$ = 0 and $m_S$ = ±1 sublevels, as seen in the ground-state spin Hamiltonian (27):

$$H_{NV} = DS_z^2 + g\mu_B \vec{B}\vec{S} + \vec{S}\overline{A}\vec{I}. \tag{1}$$

Here $\vec{S}$ is the electronic spin operator, $g$ = 2.00 is the electron g factor, $\mu_B$ is the Bohr magneton, $\vec{B}$ is the applied magnetic field, $\overline{A}$ is the hyperfine tensor, and $\vec{I}$ is the nitrogen nuclear spin operator ($I$ = 1). Additional considerations regarding Eq. (1), such as the influence of strain and electric fields, are discussed in the SI Appendix. The crystal field splitting, approximately 2.870 GHz at room temperature, exhibits shifts of -74 kHz/K due to thermal expansion (20) and vibronic interactions (28). Since the $D$ term is second order in $S_z$, ground-state level shifts due to changes in temperature can be distinguished from Zeeman shifts (Fig. 1B). This point has recently been used to perform NV center magnetometry that is insensitive to temperature drift (29). Here we take the opposite approach to perform thermometry that is robust against the NV center's interactions with its magnetic environment. Our approach is to resonantly manipulate the spin such that at specific times the relative phase of the spin states in the laboratory frame becomes $e^{-iDt}$, where $t$ is the spin's total free evolution time. As this phase is independent of the magnetic field (up to fluctuations on the timescale of the inter-pulse delay, τ), it is protected from the low-frequency magnetic fluctuations that limit the coherence to the inhomogeneous spin lifetime ($T_2^*$). Similarly, the phase is independent of the nitrogen nuclear spin state, thus eliminating interference effects observed in conventional DC sensing techniques based on Ramsey measurements (Fig. 1C). This relative phase is then converted into a population difference of the spin sublevels and measured through the fluorescence intensity ($I_{PL}$). The fluorescence is ~30% greater for the $m_S = 0$ state than for the $m_S = \pm 1$ states, and varies linearly with the $m_S = 0$ spin population, providing a direct measurement of the NV center spin. This measurement sequence is repeated until the desired level of $I_{PL}$ is reached. As a function of $t$, the $I_{PL}$ signal oscillates between $I_{PL}(m_S = 0)$ and $I_{PL}(m_S = \pm 1)$ with a frequency given by $D - \Omega_{REF}$, where $\Omega_{REF}$ corresponds to the microwave carrier frequencies used for spin manipulation. Small thermally driven changes in $D$ (as well as the value of $D$ itself) can therefore be measured through large changes in the NV center's fluorescence.

**Fluorescence Thermometry in Finite Magnetic Fields**

We first demonstrate 'thermal' analogs to conventional Hahn echo and CPMG-N pulse sequences (30) in finite magnetic fields, where both the $m_S = 0$ to $m_S = -1$ and $m_S = 0$ to $m_S = +1$ transitions are resonantly addressed. After optically initializing the spin into the $m_S = 0$ sublevel with a ~2 μs laser pulse, we apply one of the pulse sequences shown in Figure 2A, referred to here as the thermal echo (TE) and thermal CPMG-N (TCPMG-N). These pulse sequences follow the same general methodology: the spin is first initialized into a superposition of two of its three eigenstates and then undergoes periods of free evolution punctuated by operations of the form $\pi_{+1}\pi_{-1}\pi_{+1}$ or $\pi_{-1}\pi_{+1}\pi_{-1}$, where the subscripts indicate the spin transition being addressed; e.g., $\pi_{-1}$ acts on the $m_S = 0$ to $m_S = -1$ transition. These operations interchange the spin populations in the $m_S = +1$ and $m_S = -1$ states, thus inverting the NV center spin state. As a result of the inversions, the $m_S = +1$ and $m_S = -1$ states acquire a total phase $e^{-iDt}$, which is independent of static magnetic fields and low-frequency magnetic noise. The final microwave pulse converts this phase into a population of the $m_S = 0$ state, which is measured optically. The key difference between the TE and TCPMG-N sequences is that the TCPMG-N sequences invert the spin more frequently (2N times), and thus counteract higher-frequency magnetic noise. Note that the final $\pi_{-1}\pi_{+1}\pi_{-1}$ operation in the TE sequence compensates for undesired phases acquired during the duration of the π pulses and the interpulse delays (see SI Appendix for more details). The TCPMG-N sequences also include two symmetrized $\pi_{-1}\pi_{+1}\pi_{-1}$ operations per period, compensating for these undesired phase accumulations and providing better protection from noise (31).

We apply these pulse sequences to a single NV center in a diamond with natural $^{13}$C abundance (1.1%) in a magnetic field ($B_z$ = 30 G). In these materials the $^{13}$C nuclear spins ($I$ = ½) are the primary source of decoherence, limiting $T_2^*$ to a few microseconds. Conventional Hahn echo measurements can extend the coherence to hundreds of microseconds, but are limited by the incoherent precession of the $^{13}$C nuclei, which causes the coherence to collapse and revive at the $^{13}$C Larmor frequency (32). Previous studies demonstrated that dynamical decoupling sequences, such as CPMG-N, can counteract the $^{13}$C precession to maintain coherence over significantly enhanced timescales (23), and here we demonstrate that this coherence can be accessed for DC thermal sensing. Our focus is on extending the timescale of the first coherence collapse, since our three-level pulse protocols lead to three incommensurate $^{13}$C precession frequencies that do not produce the coherence revivals observed in the two-level Hahn echo (33). Figure 2B shows the Hahn echo coherence, with the width of the gray shaded region corresponding to the spin's $T_2^*$. The TE sequence achieves a coherence time similar to that of the Hahn echo sequence, approximately four times greater than $T_2^*$ (Fig. 2C). The TCPMG (Fig. 2D) and TCPMG-2 (Fig. 2E) sequences extend the spin coherence even further in time. The coherence time reaches 17.6 µs for TCPMG-2, corresponding to a 9x improvement in coherence over $T_2^*$ and a 3x improvement in thermal sensitivity ($\eta$) as compared to a Ramsey measurement. Using Eq. (4) described in the Methods, which is based on monitoring thermally driven changes in $I_{PL}$ at the free evolution time which mutually optimizes the coherence and accumulated phase, we infer that for these experimental parameters $\eta$ is 54 ± 1 mK Hz$^{-1/2}$. Combining our observed coherence with higher photon collection efficiencies achieved in scanning diamond magnetometers (34) could improve $\eta$ to ~15 mK Hz$^{-1/2}$.

Our measurements show good agreement with the results of numerical modeling for the spin dynamics of an NV center in a $^{13}$C spin bath subjected to the TE and TCPMG-N sequences (solid green lines in Fig. 2B-E; see the Methods for further details). We note that the difference between the modeling results and the experimental data is larger for the sequences involving more pulses, suggesting that they are caused by pulse imperfections. Interestingly, the simulations for TCPMG-2 (Fig. 2E) show a distinct feature near 25 µs, when the coherence collapses and then revives with a 180° shift in phase in the oscillations. We have observed similar behavior on other NV centers in our diamond sample (see SI Appendix) and tentatively attribute these features to coupling to the most proximal $^{13}$C nuclei (32). The overall correspondence between the coherence observed in the measurements and in the numerical simulations supports the assertion that the $^{13}$C nuclear spins are a primary limiting factor for these measurements. These observations suggest that the $^{13}$C dynamics could be further suppressed to produce improved thermal sensitivities not only through the application of higher order TCPMG-N sequences, but also by reducing the magnetic field to lower the $^{13}$C precession frequency.

**Fluorescence Thermometry Near Zero Magnetic Field**

Near zero magnetic field, the bandwidth of the resonant microwave pulses exceeds the spectral width of the $m_S$ = +1 and $m_S$ = -1 resonances, and the thermal echo can be performed with a single microwave source (Fig. 3A). We consider near-zero magnetic fields ($B_z$ ~ 0.5 G) to mitigate mixing of the $m_S$ = ±1 states induced by transverse strain (18). In this limit the TE sequence has an analogous form to pulse sequences proposed for NV center timekeeping (35), where the π pulse that inverts the spin state has a duration such that a spin initially in $m_S$ = 0 goes through a superposition of $m_S$ = ±1 and returns to $m_S$ = 0 with an opposite sign.

In this small magnetic field, the TE sequences produce longer coherence times of ~85 μs (Fig. 3B), in good agreement with numerical simulations. This improves the inferred $\eta$ to 25 ± 2 mK Hz$^{-1/2}$ for these measurements and we estimate that improved fluorescence detection techniques could improve $\eta$ to ~7 mK Hz$^{-1/2}$. To show the sensitivity of this coherence to temperature shifts, we perform a direct experimental demonstration where we increase the temperature of the diamond substrate by 0.1 K while keeping the microwave carrier frequency ($\Omega_{REF}$) fixed (Fig. 3C). The change in the accumulated phase of the $m_S$ = +1 and $m_S$ = -1 states produces a pronounced change in the $I_{PL}$ signal—the dashed blue line spanning Figure 3B and Figure 3C draws attention to when the two signals have opposite phase. As the 0.1 K shift is clearly resolved, these measurements demonstrate that NV centers could offer a significant improvement in thermal sensitivity relative to alternative nanoscale fluorescence thermometers (3, 4, 8-10). Even longer coherence times approaching ~1 ms, achieved through the application of higher order decoupling sequences or the use of isotopically purified diamond to eliminate the $^{13}$C spin bath (36), could further improve the thermal sensitivities from those discussed here.

Finally, to demonstrate the versatility of these NV center thermometry techniques, we perform a similar TE measurement at 500 K (Fig. 4). This measurement shows a coherence similar to the 300 K thermal echoes as expected from the temperature-independence of the $^{13}$C spin bath (19) and the sufficiently long spin-lattice relaxation time (~1 ms) at this temperature (37). The 500 K TE does show a reduced fluorescence contrast, predominantly due to an increased fluorescence background at this temperature. However, this measurement still achieves a comparable $\eta$ of 39 ± 6 mK Hz$^{-1/2}$ due to the larger shift in *D* (-140 kHz/K at 500 K), while enhanced photon collection efficiencies could provide an *η* of ~11 mK Hz$^{-1/2}$. The robust temperature dependence of these spin-based thermometry methods suggests they could be applied in a variety of contexts beyond intracellular sensing such as diamond-based scanning thermal microscopy (34, 38).

**Conclusions**

These results illustrate the NV center's promise for nanoscale thermometry applications by demonstrating dynamical decoupling techniques that harness the NV center's long spin coherence for thermal measurement. By utilizing this quantum degree of freedom, we achieve thermal sensitivities that approach 10 mK Hz$^{-1/2}$, demonstrating the NV center's potential to improve on existing nanoscale fluorescence thermometry techniques by an order of magnitude. While these results are obtained using single NV centers in high-quality synthetic diamond, the development of diamond nanostructures containing highly coherent NV centers (39, 40) suggests a pathway for fabricating nanodiamonds that achieve similar sensitivities. Such nanostructures could enable high precision thermometry in biological and microfluidic systems with sub-diffraction spatial resolution defined by the size of the nanodiamonds. Furthermore, scanning thermal measurement in fluids will benefit from recent advances in the three-dimensional spatial control and rotational control of optically trapped nanodiamonds containing NV centers (41, 42). These compelling applications suggest the potential impact of nanoscale thermometers that combine the solid-state quantum coherence of NV centers demonstrated here with the biocompatibility, hardness, and high thermal conductivity of diamond.

**Methods**

**Temperature Sensitivity Estimates –** The measurements presented here are principally concerned with determining the coherence of the TE and TCPMG-N pulse sequences. Thus, for our measurements we detune our microwave carrier frequencies from *D* by an amount sufficient to induce oscillations in $I_{PL}$ as a function of *t*, the free evolution time. We fit the finite field results to functions of the form

$$I_{\text{PL}} = a\, e^{-\left(\frac{t}{\tau_{1/e}}\right)^m} \cos(\omega t + \varphi) + b, \tag{2}$$

where $\tau_{1/e}$ is the 1/e decay time and *a*, *m*, *ω*, *φ* and *b* are free parameters. We fit the near-zero-field data to functions of the form,

$$I_{\text{PL}} = a\, e^{-\left(\frac{t}{\tau_{1/e}}\right)^m} \left(\frac{1}{4} + \cos(\omega t + \varphi)\right) + b, \quad (3)$$

to account for the fact that the π/4 pulse initializes the spin into $\frac{-i}{2}|1\rangle + \frac{-i}{2}|-1\rangle + \frac{1}{\sqrt{2}}|0\rangle$, such that at long times $I_{\text{PL}}$ decays to $\frac{3}{8}I_{\text{PL}}(m_S = 0) + \frac{5}{8}I_{\text{PL}}(m_S = \pm 1)$. From our measurements we also extract the number of photons per measurement shot for the bright and dark spin states ($p_0$ and $p_1$, respectively). We use this information to quantify $\eta$, defined as a signal to noise ratio of one, using an expression analogous to those considered for NV center magnetic sensing (16, 21):

$$\eta = \sqrt{\frac{2(p_0 + p_1)}{(p_0 - p_1)^2}} \frac{1}{2\pi \frac{dD}{dT} e^{-\left(\frac{t}{\tau_{1/e}}\right)^m} \sqrt{t}}, \quad (4)$$

where the $\eta$ values quoted in the main text correspond to the maximum value of $e^{-\left(\frac{t}{\tau_{1/e}}\right)^m} \sqrt{t}$. In our experiments we obtain $p_0$ and $p_1$ values of ~0.007 and ~0.005, respectively; for the case of optimized photon collection efficiencies we assume values of ~0.09 and ~0.065 (26, 34). In addition to assuming that measurements are performed by monitoring $I_{\text{PL}}$ at the optimal $t$, Eq. (4) also assumes the limit of small detunings between $D$ and $\Omega_{\text{REF}}$ and that the initialization and readout microwave pulses are phase shifted by 90° so that a linear relationship between the relative phase of the spin states and the measured $I_{\text{PL}}$ is achieved. For an extensive discussion of these considerations in the context of NV center magnetometry, see Ref. (21).

**Numerical Modeling –** The positions of the $^{13}$C nuclear spins were randomly generated at sites inside a 70 x 70 x 70 atoms$^3$ diamond lattice with an abundance of 1.08 %. The dipolar interactions between the NV center spin and the $^{13}$C spins were considered, while the dipolar couplings between the nuclear spins were neglected (a good approximation for our relevant experimental times of 50 – 100 μs). The full quantum dynamics of the NV center electronic spin ($S = 1$) and the 1200 most strongly coupled $^{13}$C spins ($I = ½$) were modeled. The quantitative details of the NV center spin dynamics depend on the specific positions of the $^{13}$C nuclei around the NV center, and can be strongly influenced by the most proximal $^{13}$C (32, 43). However, the overall shapes of the TE and TCPMG-N signals, as well as the typical timescales, are relatively consistent between different nuclear configurations. A further discussion of the NV center decoherence caused by the $^{13}$C spin bath is provided in the SI Appendix.

**Acknowledgements**

We acknowledge financial support from the AFOSR and DARPA. Work at Ames Laboratory was supported by the Department of Energy, Basic Energy Sciences under contract number DE-AC02-07CH11358. The authors thank F.J. Heremans for technical assistance and B. B. Buckley, A.L. Falk, C.G. Yale, and A.L. Yeats for helpful discussion.

**Competing Financial Interests**

The authors declare no competing financial interests.

**Author contributions**

All authors designed the research and analyzed the data. D.M.T., C.F.C., and D.J.C processed the devices and performed the measurements. V.V.D. performed the numerical modeling. All authors helped in writing the manuscript.

This article contains supporting information online at:

*Note added* – After submission of this manuscript two complementary preprints discussing NV center thermometry were posted online (44, 45).

**Figure Legends**

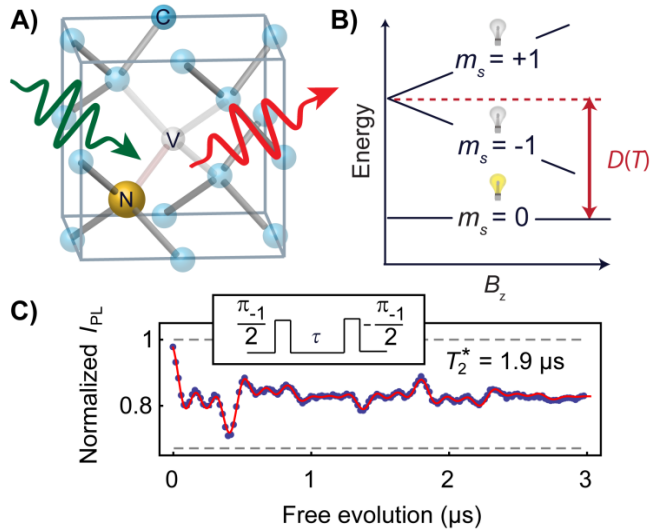

**Figure 1: The nitrogen vacancy center in diamond**

**A**, Depiction of a nitrogen vacancy (NV) center in the diamond lattice. The wavy green arrow represents the 532 nm laser used for optical excitation and the wavy red arrow represents the phonon-broadened fluorescence used to measure the spin state. **B**, Fine structure of the NV center ground state as a function of axial magnetic field. The light bulbs represent the relative fluorescence difference for the $m_s$

= 0 and $m_S = \pm 1$ states. Temperature changes shift the crystal field splitting (D), whereas magnetic fields (B) split the $m_S = \pm 1$ sublevels. This difference enables dynamical decoupling pulse sequences that move the spin between all three states to selectively measure temperature shifts and mitigate magnetic noise. **C**, Ramsey measurement performed on the $m_S = 0$ to $m_S = -1$ transition ($B_z = 40$ G). The inset illustrates the pulse sequence. The short inhomogeneous spin lifetime ($T_2^*$) limits the sensitivity of conventional NV center thermometry. The uncertainties in $I_{PL}$, estimated from the photon shot noise, are ~0.003. The microwave carrier frequency was detuned from the $m_I = 0$ hyperfine resonance by ~3.5 MHz to induce oscillations in $I_{PL}$. The fluorescence signal exhibits a beating caused by the three hyperfine resonances and weak coupling to a nearby $^{13}C$ spin. The gray dashed lines show the fluorescence intensity of $m_S = 0$ and $m_S = -1$ as determined by independent measurements.

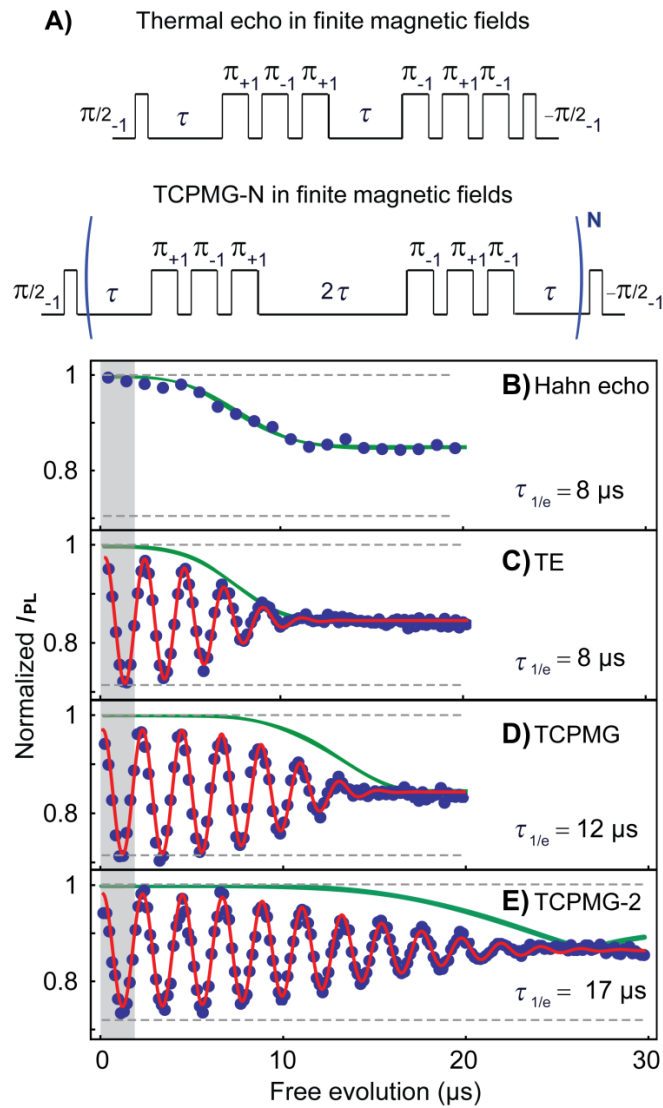

A) Thermal echo in finite magnetic fields

TCPMG-N in finite magnetic fields

B) Hahn echo — $\tau_{1/e} = 8$ µs

C) TE — $\tau_{1/e} = 8$ µs

D) TCPMG — $\tau_{1/e} = 12$ µs

E) TCPMG-2 — $\tau_{1/e} = 17$ µs

**Figure 2: Thermal echo and thermal CPMG-N sequences in finite magnetic fields**

**A**, Diagram showing the thermal echo (TE) and thermal CPMG-N (TCPMG-N) pulse sequences in finite magnetic fields. Here a π pulse has a duration such that it will invert the spin population between the two sublevels that are resonantly addressed. The subscripts indicate the spin transition being addressed. **B**, Hahn echo measurement ($m_S = 0$ to $m_S = -1$ transition) at $B = 30$ G showing $I_{PL}$ as a function of the free evolution time ($2\tau$ for the Hahn echo and TE, $4\tau$ for TCPMG and $8\tau$ for TCPMG-2). The measurement demonstrates the coherence collapse caused by the incoherent precession of the $^{13}C$ spin bath. The width of the shaded gray region represents $T_2^*$. Panels **C**, **D**, and **E**, show TE, TCPMG, and TCPMG-2 measurements, respectively, performed at 297.00 K (see SI Appendix). The uncertainties in $I_{PL}$, estimated from the photon shot noise, are ~0.005. In order to induce oscillations in $I_{PL}$ to clearly observe the signal envelope, the average microwave carrier frequency ($\Omega_{REF}$) was detuned from $D$ by ~0.5 MHz. The solid red lines are best fits to the data and the 1/e decay times ($\tau_{1/e}$) are noted on the plots. The observed coherences are in good agreement with numerical modeling (solid green lines). For the TCPMG-2 sequence the coherence time is 9x greater than $T_2^*$. For these experimental conditions we infer the thermal sensitivity ($\eta$) is $54 \pm 1$ mK Hz$^{-1/2}$. Enhancements in the photon collection efficiency (see the Methods) could improve $\eta$ to ~15 mK Hz$^{-1/2}$.

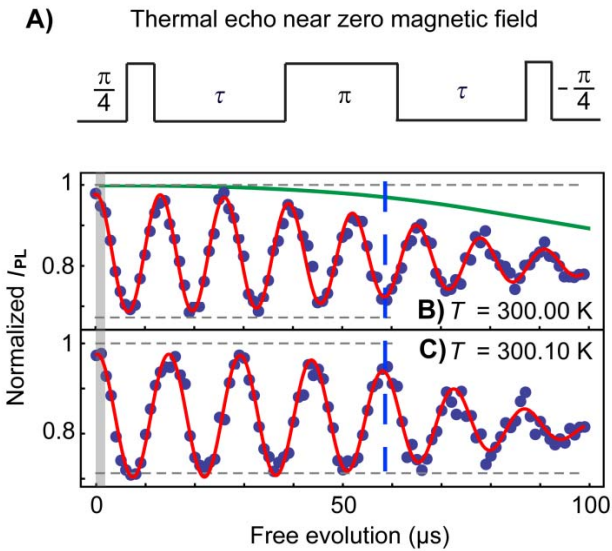

**Figure 3: Thermal echo sequence near zero magnetic field**

**A**, Diagram showing the thermal echo pulse sequence in zero magnetic field when the frequency separation of the $m_S = \pm 1$ sublevels is less than the bandwidth of the microwave pulse used to address the transitions (25 MHz). Here a π pulse has a duration such that the spin, initialized into $m_S = 0$, goes through a superposition of the $m_S = \pm 1$ states and returns to $m_S = 0$ with an opposite sign. **B**, Thermal echo measurement at 300.00 K. The microwave carrier frequency ($\Omega_{REF}$, 2.87016 GHz) was detuned from $D$ by ~75 kHz in order to induce oscillations in the signal to observe the coherence envelope. The solid green line shows the coherence predicted by the numerical modeling. The width of the shaded gray region represents $T_2^*$. **C**, The same measurement as in **B**, but with the sample temperature at 300.10 K. The shift in $D$ results in a pronounced difference in the signal frequency leading to large $I_{PL}$ differences at long times (dashed blue line). For both measurements $\tau_{1/e}$ is ~85 μs, and the uncertainties in $I_{PL}$, estimated from the photon shot noise, are ~0.02. For reference, for these data the measurement time per point was 50 seconds. From the observed coherence times we infer $\eta$ is 25 ± 2 mK Hz$^{-1/2}$ for these experimental conditions; we estimate that enhanced photon collection efficiencies could improve $\eta$ to ~7 mK Hz$^{-1/2}$. The frequency difference for the oscillations in **B** and **C**, inferred from the fits, is 8 ± 2 kHz, in good agreement the expected value ~7.4 kHz (20).

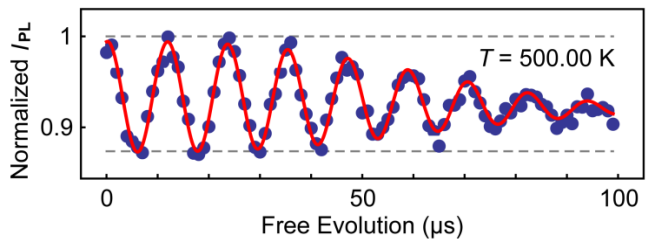

**Figure 4: Thermal echo at 500 K**

A similar measurement to those presented in Figure 3 performed at 500.00 K. Here $\Omega_{REF}$, 2.84818 GHz, has been adjusted to compensate for the large shift in $D$ such that a detuning of ~80 kHz is achieved. The uncertainties in $I_{PL}$, estimated from the photon shot noise, are ~0.007. The measurement shows a reduced $I_{PL}$ contrast between the spin states, primarily due to an increased fluorescence background on this sample at elevated temperatures. However, the larger thermal shifts in $D$ (-140 kHz/K at 500 K, Ref. (19)), largely compensate for this reduction to produce an inferred $\eta$ of 39 ± 6 mK Hz$^{-1/2}$ for this measurement. The projected $\eta$ for enhanced photon collection efficiencies is ~11 mK Hz$^{-1/2}$.